\documentclass[twocolumn,aps,prb,superscriptaddress,showpacs]{revtex4}

\usepackage{amsmath,amssymb}
\usepackage{graphicx}

\newcommand\F{{\scriptscriptstyle \rm F}}

\begin{document}

\title{Influence of a Random Telegraph Process on the Transport through a
Point Contact}

\author{Fabian Hassler}
\affiliation{Theoretische Physik,
ETH Zurich, CH-8093 Zurich, Switzerland}

\author{Gordey B.\ Lesovik}
\affiliation{L.D.\ Landau Institute for Theoretical Physics RAS,
117940 Moscow, Russia}

\author{Gianni Blatter}
\affiliation{Theoretische Physik,
ETH Zurich, CH-8093 Zurich, Switzerland}

\date{\today}

\begin{abstract}
	We describe the transport properties of a point contact under the
	influence of a classical two-level fluctuator. We employ a transfer
	matrix formalism allowing us to calculate arbitrary correlation
	functions of the stochastic process by mapping them on matrix
	products. The result is used to obtain the generating function of the
	full counting statistics of a classical point contact subject to a
	classical fluctuator, including extensions to a pair of two-level
	fluctuators as well as to a quantum point contact. We show that the
	noise in the quantum point contact is a sum of the (quantum)
	partitioning noise and the (classical) noise due to the two-level
	fluctuator. As a side result, we obtain the full counting statistics
	of a quantum point contact with time-dependent transmission
	probabilities.
\end{abstract}


\maketitle

\section{Introduction}

The impact of mobile impurities on transport through a quantum conductor
attracted a lot of attention soon after it was realized that the
conductance of a dirty coherent sample is sensitive to the position of
a single impurity; this discovery formed the basis for the explanation
of flicker noise as it appears due to the presence of bistable mobile
impurities.\cite{altshuler:89,feng:86} A detailed characterization of
charge transport through a quantum conductor is provided by the full
counting statistics (FCS).\cite{levitov:96} During the past decade, this
description has been applied to numerous systems \cite{nazarov_blanter} and
a first attempt to describe the influence of a two-level fluctuator on the
FCS of quantum transport has been given in Ref.\ \onlinecite{lesovik:94b}.
In the present paper, we calculate the full counting statistics of charge
transport through a (classical or quantum) point contact coupled to a
classical two-level fluctuator; we go beyond previous studies by considering
the combined effects of the presence of one or many mobile impurities,
as well as quantum-partitioning and the Fermi statistics on the FCS. Also,
we reconsider carefully the case when partitioning is neglected and correct
previous findings which are flawed when calculating the fourth or higher
cumulants.

The influence of a fluctuating (uncontrollable) environment on a
(controllable) device is a generic problem \cite{feynman:63} and our work is
related to other studies, e.g., the transport statistics through a quantum dot
in the Coulomb-blockade regime \cite{bagrets:03,gustavsson:06} or the effect
of a bistability on the transport through a quantum point
contact\cite{jordan:04,sukhorukov:07}. Another example is the dephasing of
qubits due to a classical \cite{falci:03} or quantum \cite{abel:08} two-level
system, or the studies on $1/f$-noise originating from telegraph noise due to
classical \cite{schriefl:05,schriefl:06} or quantum
\cite{paladino:02,falci:03,galperin:06,bergli:09} two-level fluctuators.  The
two problems, full counting statistics of charge transport and dephasing of a
quantum system (qubits) are related through the equivalence of fidelity
\cite{peres:84} and full counting statistics \cite{levitov:96}, as has been
pointed out recently.\cite{lesovik:06}

In our analysis below, we describe the time evolution of the two-level
fluctuator by rate equations which can be solved explicitly. We then study
the full counting statistics of a wire with a conductance depending on
the state of the fluctuator. The fluctuator induces noise in the transport
current through the wire which we evaluate using a mapping of correlation
functions on matrix products. Using this mapping, we are able to calculate
the full counting statistics of a classical wire coupled to a two-level
fluctuator.  Furthermore, we discuss the situation of a second (independent)
fluctuator and show that a nonlinear interaction with the wire can lead to
correlations in the noise even though the fluctuators evolve independent of
each other.  Finally, we apply our method to the case of a quantum wire which
exhibits intrinsic partitioning (shot) noise.\cite{khlus:87,lesovik:89} We
derive a formula which incorporates both classical- (due to the two-level
fluctuator) and quantum- (due to the point contact) noise. Thereby, we
give an explicit expression for the full counting statistics of a quantum
point contact whose transmission probabilities change with time. As
two-level fluctuators seem to be a major obstacle for achieving solid
state implementations of qubits with long coherence times, being able to
characterize the influence of a two-level fluctuator on transport through
a point contact offers the possibility to learn about the fluctuating
environment by measuring the full counting statistics through a nearby
quantum point contact. The (partial) overlap of our results with previous
work\cite{bagrets:03,falci:03,jordan:04} will be discussed below.
\begin{figure}[t]
  \centering
  \includegraphics{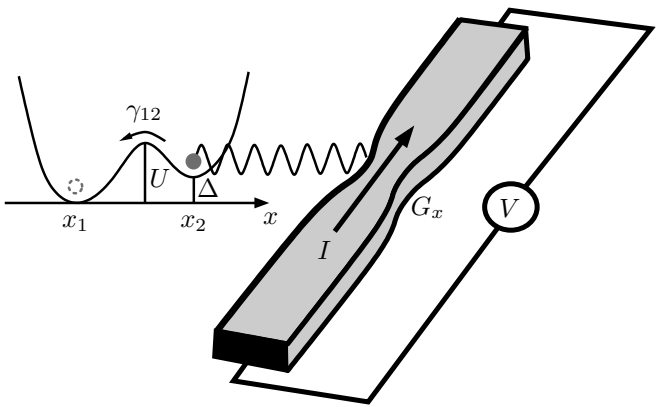}
  \caption{%
  Sketch of the setup: A two-level system fluctuating incoherently between
  states $x_{1,2}$ with rates $\gamma_{12}$ and $\gamma_{21}$ is coupled
  to a wire with a constriction. The conductance $G$ of the wire changes
  according to the state of the fluctuator, inducing noise in the current
  flowing through the device.
  }\label{fig:two_level}
\end{figure}

\section{Single Two-Level Fluctuator}\label{sec:single}

Consider a classical particle which can be trapped in an external
potential at two positions denoted by $x_1$ and $x_2$, cf.\
Fig.~\ref{fig:two_level}. The potential is characterized by the energies
$E_{1,2}$ associated with the two valleys, which are shifted by the amount
$\Delta = E_2 -E_1$ with respect to each other, and the height of the
barrier $U$. We assume the dynamics to be given by thermally activated
hopping over the barrier.\cite{kramers:40} The particle performs a random
(Brownian) motion where the probabilities $P_{1,2}$ to be in either valley
obey the rate equations
\begin{align}\label{eq:rate_eq}
  \dot P_1(t) &= -\gamma_{21} P_1(t) + \gamma_{12} P_2(t), \nonumber\\
  \dot P_2(t) &= \phantom{-}\gamma_{21} P_1(t) - \gamma_{12} P_2(t),
\end{align}
where the rate to escape from $x_1$ to $x_2$ is given by $\gamma_{21}=
\gamma_{1\to2} \propto e^{-U/\vartheta}$ and the reverse process from
$x_2$ to $x_1$ is governed by the rate $\gamma_{12}=\gamma_{2\to1}
\propto e^{(\Delta-U)/\vartheta}$, with $\vartheta$ the temperature. In
equilibrium, the probabilities $P_{1,2}^\text{eq}$ are such as to obey
the balance equation $dN_{21} = dN_{12}$ which equates the number of
particles $d N_{21} = \gamma_{21} P_1^\text{eq} dt$ going from $x_1$
to $x_2$ during the time $dt$, with the number of particles $dN_{12} =
\gamma_{12} P_2^\text{eq} dt$ going the opposite way. The equilibrium
probabilities therefore satisfy the Gibbs weight
\begin{equation}\label{eq:gibbs}
  P_2^\text{eq}/P_1^\text{eq} = e^{-\Delta/\vartheta}
\end{equation}
and together with the probability conservation $P_1^\text{eq} + P_2^\text{eq}
= 1$, we obtain
\begin{align}\label{eq:equilibrium}
  P_1^\text{eq} &= \gamma_{12}/\Gamma, &
  P_2^\text{eq} &= \gamma_{21}/\Gamma,
\end{align}
where we have introduced the total rate $\Gamma=\gamma_{12}+\gamma_{21}$.

Introducing the vector $\mathbf{P}(t)$ with components $P_{1,2}(t)$, the
rate equation (\ref{eq:rate_eq}) can be written as $\dot{\mathbf{P}}(t)
= - \mathsf{h} \, \mathbf{P}(t)$, with the Fokker-Planck
Hamiltonian\cite{zinn-justin} given by
\begin{equation}\label{eq:hamiltonian}
  \mathsf{h} =
  \begin{pmatrix}
    \gamma_{21}   &  - \gamma_{12} \\
    - \gamma_{21} &  \gamma_{12}
  \end{pmatrix}.
\end{equation}
Note that the Hamiltonian $\mathsf{h}$ is not Hermitian (left and right
eigenvalues are not simply adjoint to each other). Nevertheless, its
eigenvalues are real (and even positive). To make this point clear, we write
the Hamiltonian $\mathsf{h}$ in a new basis $\mathsf{h}'= \mathsf{s}^{-1}
\mathsf{h} \mathsf{s}$ using the transformation matrix $\mathsf{s} =
\text{diag}(\sqrt{P^\text{eq}_1},\sqrt{P^\text{eq}_2})$, such that
\begin{equation}\label{eq:hp}
  \mathsf{h}' = 
  \begin{pmatrix}
    \gamma_{21} & - \Gamma \sqrt{P^\text{eq}_1 P^\text{eq}_2}\\
     - \Gamma \sqrt{P^\text{eq}_1 P^\text{eq}_2} & \gamma_{12}
  \end{pmatrix}.
\end{equation}
It is now visible that the matrix $\mathsf{h}'$ (and therefore also the
matrix $\mathsf{h}$) has real (the matrix is symmetric) and positive
(the determinant and the trace of the matrix are positive) eigenvalues.
The evolution conserves the probability as $\partial_t [P_1(t) + P_2(t)]
=0$, as is evident from Eq.~(\ref{eq:rate_eq}). This relation implies that
$(1,1) \cdot \dot{\mathbf{P}}(t) =0$ and it follows that $\langle 0 |=
(1,1) $ is a left eigenvector of the Hamiltonian $\mathsf{h}$ to the
eigenvalue $0$. To every left eigenvector there exists a corresponding
right eigenvector with the same eigenvalue which we will denote by $|
0 \rangle$. The right eigenvector to the eigenvalue $0$ is given by the
equilibrium distribution $| 0 \rangle = \mathbf{P}^\text{eq}$, $\mathsf{h}
| 0 \rangle = 0$. The second eigenvalue is given by $\Gamma$ with the
corresponding right [left] eigenvectors assuming the form $| \Gamma \rangle =
(1,-1)^T$ [$ \langle \Gamma | = (P_2^\text{eq}, -P_1^\text{eq})$]; note
that the eigenvectors are normalized such that $\langle a | b \rangle
=\delta_{ab}$ and $\sum_a |a \rangle \langle a| = \openone_2$ with $a,b
\in \{0,\Gamma\}$. Using the eigenbasis of $\mathsf{h}$, it is possible
to compute the evolution operator $\mathsf{P}(t>0) = \exp(- \mathsf{h} t)=
\sum_a e^{-a t} | a \rangle \langle a |$. The matrix element $\mathsf{P}_{mn}
(t)$ denotes the conditional probability for the particle to be transferred
from state $n$ to $m$ in the time $t$.  The evolution only depends on the
time difference as the Hamiltonian $\mathsf{h}$ is time-independent. An
explicit calculation yields the expression
\begin{equation}\label{eq:prop}
  \mathsf{P}(t) =
  \begin{pmatrix}
    P_1^\text{eq} + P_2^\text{eq} e^{-\Gamma t} & 
    P_1^\text{eq} (1 - e^{-\Gamma t}) \\
    P_2^\text{eq} (1 - e^{-\Gamma t}) &
    P_2^\text{eq} + P_1^\text{eq} e^{-\Gamma t} \\
  \end{pmatrix}
\end{equation}
for the propagator during the time $t$. As the stochastic process is
Markovian, the propagator (\ref{eq:prop}) incorporates all the information
needed in order to calculate general correlation functions of the stochastic
process,\cite{gardiner} see also below.

\section{Classical Wire}

We consider a classical wire coupled to the two-level fluctuator. We assume
the two-level system to be a charge impurity which interacts with the wire,
e.g., via Coulomb forces. The net effect of the charge impurity is to change
the conductance of the wire $G_{1,2}$ depending on the position $x(t)=
x_{1,2}$ of the two-level system. The wire is biased by a constant voltage $V$
such that the current is determined by $I_{1,2} = V G_{1,2}$. The current in
the wire jumps between $I_1$ and $I_2$ in a random way given by the dynamics
of the two-level fluctuator which we assume to be in thermal equilibrium,
i.e., in the state $|0\rangle$. The fluctuations of the two-level system induce
current noise. In our discussion, any kind of back-action of the wire on the
two-level fluctuator is neglected.

\subsection{Correlation Functions -- Mapping on Matrices}\label{sec:mapping}

With our focus on the full counting statistics, we are interested
in obtaining the moments of the charge $Q = \int_0^t dt_1 \, I(t_1)$
transmitted through the point contact during the time $t$; here, $I(t_1)$
denotes $I_{1,2}$ depending on the state $x(t_1)=x_{1,2}$ of the two-level
fluctuator at time $t_1$. As $I(t_1) = V G(t_1)$, we first concentrate
on correlation functions of $G(t)$, where statistical averages over
the stochastic process (\ref{eq:rate_eq}) will be denoted by $\langle
\cdot \rangle$. In quantum mechanics, it is well-known that correlation
functions can be evaluated either in the operator or in the path-integral
formalism.\cite{hibbs} Likewise, we have the choice to apply a stochastic
path-integral approach\cite{bagrets:03,pilgram:03a} or to use the operator
formalism. Here, we stay with the operator formalism introduced in the
previous section. Note that the propagator $\mathsf{P}_{mn} (t_2-t_1)$,
cf.\ Eq.~(\ref{eq:prop}), denotes the conditional probability (the transfer
matrix) to find the system in state $x_m$ at time $t_2$, given that it
resided in $x_n$ at time $t_1$, $\mathsf{P}_{mn} (t_2-t_1) = \langle x(t_2)
{=} x_m | x(t_1) {=} x_n \rangle$; i.e., within a path-integral formulation,
$\mathsf{P}(t)$ involves already an integration over all possible paths
between $t_1$ and $t_2$.

The average conductance $\langle G(t_1) \rangle$ is readily calculated.
For a system residing in a stationary state given by $\langle x(t_1) {=}
x_n \rangle = P^\text{eq}_n$, we obtain
\begin{equation}\label{eq:av_cond}
  G(t_1) = \sum_{n=1,2} G_n \, 
  \langle x(t_1) {=} x_n \rangle = G_1 P_1^\text{eq} +
  G_2 P_2^\text{eq}.
\end{equation}
The calculation of the conductance correlator $\langle G(t_2) G(t_1)
\rangle$ is more involved. We proceed slowly in order to motivate our
general mapping between the calculation of correlation functions and
the evaluation of matrix products of specific matrices.  In order to
calculate $\langle G(t_2) G(t_1) \rangle$, we assume first that $t_2 > t_1$;
classical correlators are symmetric so that the opposite ordering of times
reduces to the same quantity.  Using the fact that the stochastic process
is Markovian, we expand the correlation function $\langle G(t_2) G(t_1)
\rangle = \sum_{m,n} G_m \langle x(t_2) {=} x_m | x(t_1) {=} x_n \rangle\,
G_n \langle x(t_1) {=} x_n \rangle$;\cite{gardiner} this expansion can be
seen as a transfer-matrix expansion of the correlation function. We obtain
the mapping for the correlator ($t_2>t_1$)
\begin{align}\label{eq:noise}
  \langle G(t_2) G(t_1) \rangle &= \sum_{mn} G_m 
  \mathsf{P}_{mn} (t_2 - t_1) G_n
  P^\text{eq}_n \nonumber \\
      &=  \sum_{klmn} G_k \delta_{km} \mathsf{P}_{ml}(t_2-t_1)
      G_l \delta_{ln} P_n^\text{eq} \nonumber\\
      &= \langle 0 | \mathsf{G} e^{-\mathsf{h} (t_2-t_1)} \mathsf{G} | 0
      \rangle
\end{align}
with the diagonal matrix $\mathsf{G}_{mn} = G_n \delta_{mn}$. Introducing
the ``interaction representation''
\begin{equation}\label{eq:interaction}
  \mathsf{G}_\text{I}(t) = e^{\mathsf{h} t}
  \mathsf{G} e^{-\mathsf{h} t}
\end{equation}
of the matrix $\mathsf{G}$, the correlation function Eq.~(\ref{eq:noise})
can be further simplified to
\begin{equation}\label{eq:noise_mapping}
  \langle G(t_2) G(t_1) \rangle = \mathcal{T} \langle 0 |
  \mathsf{G}_\text{I} (t_2) \mathsf{G}_\text{I} (t_1) | 0 \rangle,
\end{equation}
where the time-ordering operator $\mathcal{T}$ has been included in order
to relieve the restriction $t_2>t_1$. It is easy to see that the above
derivation is not restricted to the second order correlation function, but
can be applied in the same way to higher order correlation functions. We
thus arrive at the mapping
\begin{equation}\label{eq:mapping}
  \langle G(t_N) \cdots G(t_1) \rangle = \mathcal{T} \langle 0 |
  \mathsf{G}_\text{I}(t_N) \cdots \mathsf{G}_\text{I}(t_1) | 0 \rangle,
\end{equation}
where the left hand side is a correlation function for the classical
stochastic process involving the two-level fluctuator and the right hand
side is a matrix element involving the matrices $\mathsf{G}_\text{I}(t_n)$
and the vectors $|0 \rangle = \mathsf{P}^\text{eq}$ and $\langle 0 | =
(1,1)$.

\subsection{Full Counting Statistics}

We are now in the position to calculate the generating function 
\begin{equation}\label{eq:fcs}
  \chi (\lambda) = \langle e^{i \lambda \int_0^t dt' I(t') } \rangle
\end{equation}
for the zero-frequency current-correlation functions (moments) of a classical
point contact coupled to a two-level fluctuator. The moments are obtained as
the Taylor coefficients $\langle Q^n \rangle = (-i \partial_\lambda)^n \chi
|_{\lambda=0}$.  Alternatively, the stochastic process can be characterized
by irreducible cumulants which are given by the expansion coefficient of
the logarithm of the characteristic function
\begin{equation}\label{eq:cumulants}
  \langle\langle Q^n \rangle\rangle = \Bigl(\frac{d}{i d\lambda}\Bigr)^n
  \log \chi (\lambda) \Big|_{\lambda=0}.
\end{equation}
The characteristic function $\chi (\lambda)$ can be recast in the form
\begin{equation}\label{eq:chi_mapped}
  \chi (\lambda) = \mathcal{T} 
  \langle 0 | e^{i \lambda V \int_0^t dt' \mathsf{G}_\text{I}(t') } | 0 \rangle
\end{equation}
using the mapping Eq.~(\ref{eq:mapping}). This equation can be
rewritten using the well-known mapping between the ``Schr\"odinger''
and the ``interaction'' representation,\cite{zinn-justin}
$e^{-(\mathsf{h}+\mathsf{v})t} = e^{-\mathsf{h} t} \, \mathcal{T}
\exp[-\int_0^t dt' \mathsf{v}_\text{I}(t')]$ with $\mathsf{v}_\text{I}(t)=
e^{\mathsf{h}t} \mathsf{v} e^{-\mathsf{h}t}$.  Here, we apply the relation
in the opposite direction to arrive at an expression without the awkward
time-ordering,
\begin{equation}\label{eq:chi_final}
  \chi (\lambda) = \langle 0 | e^{(- \mathsf{h} + i \lambda V \mathsf{G}) t }
  | 0 \rangle.
\end{equation}
This formula was derived before by Bagrets and Nazarov using a stochastic
path integral formulation of the problem (the Fokker-Planck Hamiltonian
$\mathsf{h}$ and the counting field $\lambda$ are denoted by $\hat{L}$
and $\lambda$ in their paper\cite{bagrets:03}). We believe that the present
approach using transfer matrices and the mapping of the interaction picture
onto the Schr\"odinger picture is more transparent. Note though, that the
counting field $\lambda$ enters differently in their work compared to ours.
Here, $\lambda$ couples to the classical current $I$ whereas Bagrets and
Nazarov discuss the transport of individual (quantum) particles such that
$\lambda$ may only enter in the combination $\exp(i \lambda)$ due to the
quantization of charge.

For further convenience, we subtract the average charge $\langle Q \rangle
= V \langle G \rangle t$ in order to obtain the reduced full counting
statistics
\begin{equation}\label{eq:chi}
  \hat\chi (\lambda) = \chi(\lambda) e^{- i \lambda \langle Q \rangle }
  \\ =\langle 0 | e^{ - \hat{\mathsf{h}} t} | 0 \rangle,
\end{equation}
with the matrix $\hat{\mathsf{h}}=\mathsf{h} - i \lambda V (\mathsf{G}
- \langle G \rangle)$. Apart from the average charge (which is zero for
$\hat \chi$), both $\log \chi$ and $\log \hat \chi$ generate the same
cumulants. The explicit calculation of the characteristic function of the
full counting statistics $\hat\chi$ involves the eigenvalues
\[
  \hat h_\pm = \frac{\Gamma}{2}\Bigl[ 1 + 
  i \lambda \Delta g  \Delta P^\text{eq} 
  \pm
  \sqrt{
  1 + 
  2 i \lambda \Delta g \Delta P^\text{eq} -
  \lambda^2 (\Delta g)^2} \Bigr]
\]
of the matrix $\hat{\mathsf{h}}$, where we have introduced the difference
in the equilibrium population $\Delta P^\text{eq} = P^\text{eq}_2 -
P^\text{eq}_1$ and the difference in the (dimensionless) conductance
$\Delta g = V(G_2 - G_1)/\Gamma$.

\subsection{Asymptotic long-time limit}

For long times $\Gamma t \gg 1$, the matrix exponential (\ref{eq:chi})
is dominated by the eigenvalue $\hat h_{-}$ of $\hat{\mathsf{h}}$ with
the smallest real part.\cite{bagrets:03,kiesslich:06,groth:06} We obtain
an explicit expression for the generating function
\begin{equation}\label{eq:fcs_large}
  \log \hat \chi_\gg (\lambda) = - \hat h_{-} \, t.
\end{equation}
All cumulants become linear in $t$, due to the fact that the autocorrelation
time in the system is given by $\Gamma^{-1}$ and every state decays to the
equilibrium state after this time. The fluctuations for $\Gamma t \gg 1$
can be seen as a sum of independent stochastic processes and the cumulant
generating function $\log \hat \chi$ becomes extensive in $t$. Interestingly,
it is possible to obtain an explicit relation for the cumulants ($n\geq 2$)
\begin{align}\label{eq:cumulant_large}
  \langle\langle Q^{n} \rangle\rangle_\gg =&
  n! \, V^n (G_1 -G_2)^n \Gamma^{1-n} t  \\ &
  \, \times \sum_{k=1}^{n-1} N_{n-1,k} (-1)^{k+1} (P_1^\text{eq})^k
  (P_2^\text{eq})^{n-k} \nonumber,
\end{align}
with the Narayana numbers $N_{n,m} = \binom{n}{m} \binom{n}{m-1} /
n$.\cite{combinatorics} The cumulants in (\ref{eq:cumulant_large})
grow factorially in magnitude with $n$ and oscillate as a function
of $\Delta P^\text{eq}$ which are generic features of high-order
cumulants.\cite{flindt:09} The first couple of cumulants assume the form
\begin{align}\label{eq:first_cum_large}
 \langle \langle Q^2 \rangle \rangle_\gg &= 
 2  P_1^\text{eq} P_2^\text{eq} \frac{V^2 (G_1 - G_2)^2 t}
 {\Gamma}, \\ 
 \langle \langle Q^3 \rangle \rangle_\gg &= 6  
 P_1^\text{eq} P_2^\text{eq}( 
   P_2^\text{eq}- P_1^\text{eq} )
 \frac{V^3 (G_1
  - G_2)^3 t}{\Gamma^2} . \nonumber
\end{align}
Note that all the odd cumulants vanish if the process is symmetric
$\gamma_{12}=\gamma_{21}$, cf.\ Eq.~(\ref{eq:equilibrium}). The cumulants
in (\ref{eq:cumulant_large}) and (\ref{eq:first_cum_large}) agree with the
result in Ref.~\onlinecite{lesovik:94b}. However, starting with the 4-th
cumulant a discrepancy arises; e.g., for the 4-th cumulant we obtain
\begin{align*}
  \langle \langle Q^4 \rangle &\rangle_\gg = \\
  &24  
  P_1^\text{eq} P_2^\text{eq}[ ( P_2^\text{eq}- P_1^\text{eq})^2
  -P_1^\text{eq} P_2^\text{eq} ]
  \frac{V^4 (G_1
  - G_2)^4 t}{\Gamma^3},
\end{align*}
which is different from the result \cite{lesovik:94b}
\[
\langle \langle Q^4 \rangle \rangle_\gg = 24  
P_1^\text{eq} P_2^\text{eq}(  P_2^\text{eq} - P_1^\text{eq} )^2
\frac{V^4 (G_1
- G_2)^4 t}{\Gamma^3}.
\]
The latter result is incorrect as it misses terms due to the implicit
assumption in Ref.~\onlinecite{lesovik:94b} that the reduced conductance
correlators $\langle \langle G(t_n) \cdots G(t_1) \rangle \rangle$
may only depend on the largest time difference $t_n-t_1$.  Even though
this hypothesis is correct for correlators up to $n=3$, it fails for
higher-order correlators.

\subsection{Short times}

For short times, $t \ll \Gamma^{-1}$ there is no evolution of the two-level
system and we can set $\mathsf{h} t =0$. The full counting statistics reads
\begin{align}\label{eq:fcs_small}
  \hat \chi_\ll (\lambda) &= \langle 0 | e^{i\lambda V (\mathsf{G} - \langle G
  \rangle) t} | 0 \rangle \\
  &= P_2^\text{eq} e^{i \lambda V (G_2 - G_1) P_1^\text{eq} t} +
  P_1^\text{eq} e^{i \lambda V (G_1 -G_2) P_2^\text{eq} t} \nonumber ,
\end{align}
with the cumulants ($n\geq 2$) given by
\begin{align}\label{eq:cumulants_small}
  \langle \langle Q^n \rangle \rangle_\ll = & V^n (G_1 - G_2)^n t^n \\
  &\times \sum_{k=1}^{n-1} E_{n-1,k-1} (-1)^{k+1} (P_1^\text{eq})^k
  (P_2^\text{eq})^{n-k}
  \nonumber,
\end{align}
where the Eulerian numbers $E_{n,m}$ are defined through $E_{n,m} =
\sum_{k=0}^m (-1)^k \binom{n+1}{k} (m+1-k)^n$.\cite{combinatorics}
In the short time limit, the cumulants grow like $\langle\langle Q^n
\rangle\rangle\propto t^n$, i.e., higher order cumulants are suppressed
at short times. The first couple of cumulants are explicitly given by
\begin{align}\label{eq:first_cum_small}
  \langle \langle Q^2 \rangle \rangle_\gg &=  P_1^\text{eq} P_2^\text{eq}
  V^2 (G_1 - G_2)^2 t^2, \\ 
  \langle \langle Q^3 \rangle \rangle_\gg &=  P_1^\text{eq} P_2^\text{eq}( 
  P_2^\text{eq}- P_1^\text{eq} )
  V^3 (G_1
  - G_2)^3 t^3. \nonumber
\end{align}

\subsection{Arbitrary times}

Expanding the matrix exponential in Eq.~(\ref{eq:chi}) in its eigenbasis,
the generator for the full counting statistics reads
\begin{equation}\label{eq:fcs_arbitrary}
  \hat \chi = \frac{\hat h_+ e^{-\hat h_- t} - \hat h_- e^{-\hat h_+ t}}
  {\hat h_+ - \hat h_-}
\end{equation}
for arbitrary times. This result has been first derived in
Ref.~\onlinecite{falci:03} in the context of dephasing of a qubit
due to the interaction with a classical two-level fluctuator, where $\lambda$
denotes the interaction of the qubit with the fluctuator. Here, we are
interested in the transport properties of a point contact characterized
by cumulants which are given by the Taylor expansion of $\log \chi$ around
$\lambda=0$; the relation between these two problems is a consequence of
the generic equivalence between full counting statistics and fidelity,
see Ref.~\onlinecite{lesovik:06}. The first couple of cumulants are given by
\begin{align}\label{eq:cumulants_arbitrary}
  \langle \langle Q^2 \rangle \rangle &=2
  P_1^\text{eq} P_2^\text{eq} \frac{V^2 (G_1-G_2)^2 
  [(\Gamma t - 1) + e^{-\Gamma t}] }{\Gamma^2},
  \nonumber\\
  \langle \langle Q^3 \rangle \rangle &= 6  P_1^\text{eq}
  P_2^\text{eq}(P_2^\text{eq}- P_1^\text{eq} ) \\
  &\quad\times\frac{V^3 (G_1-G_2)^3
  [ (\Gamma t -2) + (\Gamma t + 2) e^{-\Gamma t} ] }{\Gamma^3}. \nonumber
\end{align}

\subsection{Symmetric fluctuator}

A special situation is given when the two-level fluctuator is symmetric,
$\Delta=0$, i.e., $P_1^\text{eq}= P_2^\text{eq}= 1/2$. Then the characteristic
function assumes the simple form
\begin{equation}\label{eq:fcs_large_sym}
  \log \chi_\gg (\lambda) =  \frac{\Gamma t}{2} \Bigl[\sqrt{1- \lambda^2
  V^2 (G_1 -G_2)^2/\Gamma^2} -1 \Bigr] 
\end{equation}
for long times. In the short time limit, the generating function
\begin{equation}\label{eq:fcs_small_sym}
  \chi_\ll(\lambda) = \cos[ \lambda V (G_1 -G_2)\, t/2 ]
\end{equation}
becomes periodic. In both cases, due to the symmetry of the states $x_1$
and $x_2$, only the even cumulants are nonvanishing.

\section{A Pair of Two-Level Fluctuators}

Needless to say, the mapping of Sec.~\ref{sec:mapping} is not restricted to
a single two-level fluctuator. It can be generalized to an arbitrary number
of states whose dynamics is governed by classical rate equations described
by a Fokker-Planck Hamiltonian $\mathsf{h}$. To illustrate this concept,
the example of two classical, uncorrelated two-level fluctuators coupled
to a wire is discussed in the following. We restrict ourselves to the case
where the dynamics of the two-level systems is completely independent of each
other such that $\mathsf{h}= \mathsf{h}^\alpha + \mathsf{h}^\beta$; here and
in the following, we denote quantities involving only the first (second)
fluctuator with a superscript $\alpha(\beta)$, e.g., $\mathsf{h}^\alpha =
\mathsf{h}^\alpha \otimes \openone^\beta$.  Written explicitly in the basis
$\{ |1\rangle^\alpha \otimes |1\rangle^\beta, |2\rangle^\alpha \otimes
|1\rangle^\beta, |1\rangle^\alpha \otimes |2\rangle^\beta, |2\rangle^\alpha
\otimes |2\rangle^\beta\}$, the Fokker-Planck Hamiltonian reads
\begin{equation}\label{eq:h_2}
  \mathsf{h} = 
  \begin{pmatrix}
    \gamma_{21}^\alpha + \gamma_{21}^\beta & -\gamma_{12}^\alpha &
    - \gamma_{12}^\beta & 0 \\
    - \gamma_{21}^\alpha & \gamma_{12}^\alpha + \gamma_{21}^\beta &
    0 & - \gamma_{12}^\beta \\
    - \gamma_{21}^\beta & 0 &
    \gamma_{21}^\alpha + \gamma_{12}^\beta & - \gamma_{12}^\alpha \\
    0 & - \gamma_{21}^\beta & 
    - \gamma_{21}^\alpha & \gamma_{12}^\alpha + \gamma_{12}^\beta
  \end{pmatrix}.
\end{equation}
One is tempted to think that the independent dynamics of the two subsystems
would lead to independent statistics, such that the characteristic function
of the full counting statistics is given by the product of the individual
characteristic functions. Indeed, this is the generic case for two-level
fluctuators coupling to a qubit, where the combined effect leads to $1/f$
noise.\cite{paladino:02,falci:03,schriefl:05,galperin:06,bergli:09} However,
here, this argument is only valid when the interaction with the wire is
``linear'' such that the effects of the individual subsystems simply add
up, in formula $\mathsf{G}= \mathsf{G}^\alpha + \mathsf{G}^\beta$. Having
the model of the Fig.~\ref{fig:two_level} in mind, this assumption is
incorrect as a (quantum) point contact does not react linearly on changes
in the gate potential and, therefore, the noise of individual fluctuators
does not simply add up. In the following, we first treat the (simple) case
of linear interaction and then comment on the correlation which arises in
the general case by applying perturbation theory in the nonlinearity.

Introducing the reference conductance $G_0 = G_0^\alpha + G_0^\beta$ as well
as the induced changes $\Delta G^x = G_1^x- G_0^x$ due to the fluctuator
$x=\alpha,\beta$, the conductance matrix for linear interaction is given by
\begin{equation}\label{eq:g_2}
  \mathsf{G} = G_0 \openone_4 + \text{diag} ( 0, \Delta G^\alpha, \Delta
  G^\beta, \Delta G^\alpha + \Delta G^\beta );
\end{equation}
the increase of conductance in the case when both fluctuators are in state
$x_2$ is the sum of the respective increases when only one of the fluctuators
is in state $x_2$, that is to say, the effects of the two fluctuators
simply add up. In this case, the characteristic function assumes the
form $\chi(\lambda) = \chi^\alpha(\lambda) \chi^\beta(\lambda)$ and the
cumulants $\langle\langle Q^n\rangle\rangle$ become a sum of cumulants
generated by the two individual subsystems.

In the case of a general (diagonal) matrix $\mathsf{G}$, the solution of the
problem involves the determination of the roots of a polynomial of fourth
degree and the characteristic function does not separate, even though the
time evolution of the two fluctuators is completely independent of each
other. To be more explicit, we want to show how a small perturbation
$\Delta G \ll \Delta G^\alpha + \Delta G^\beta$ in $\mathsf{G}_{44}$
destroying the linearity (additivity) leads to correlations which can be
arbitrary large for long times. We define $\chi^\text{corr}(\lambda) =
\chi(\lambda)/\chi^\alpha(\lambda) \chi^\beta(\lambda)$ as the part of
the characteristic function which describes the correlation between the
action of the individual subsystems. In the long-time limit, to first
order in $\Delta G$, we can apply standard perturbation theory to find the
correction to the lowest eigenvalue of Eq.~(\ref{eq:h_2}). The cumulant
generating function for the correlation is given by
\begin{multline}\label{eq:cum_gen_2}
  \log \chi^\text{corr}_\gg (\lambda) = \\
  i \lambda V \Delta G t
  \!\! \prod_{x=\alpha,\beta} \!\!\!  \frac{ (\Gamma^x + i \lambda V \Delta
  G^x) P_2^{\text{eq},x} - \hat h_-^x}
  {\hat h_+^x - \hat h_-^x} .
\end{multline}
The average transmitted charge changes according to
\begin{equation}\label{eq:average_charge_2}
  \Delta\!\langle Q \rangle = P_2^{\text{eq},\alpha} P_2^{\text{eq},\beta}
  V \Delta G \, t, 
\end{equation}
with $P_2^{\text{eq},\alpha} P_2^{\text{eq},\beta}$ the probability to be in
the state $|2\rangle^\alpha \otimes |2\rangle^\beta$ and $V \Delta G$ the
change in the current. Less trivial, the correlation contribution to the
noise
\begin{equation}\label{eq:noise_2}
  \Delta\!\langle\langle Q^2 \rangle\rangle = 4 V \Delta\! \langle Q \rangle 
  \Biggl[ \frac{P_1^{\text{eq},\alpha}} {\Gamma^\alpha} 
  \Delta G^\alpha
  +  \frac{P_1^{\text{eq},\beta}} {\Gamma^\beta} 
  \Delta G^\beta \Biggr]
\end{equation}
depends both on $\Delta G^\alpha$ and $\Delta G^\beta$.

\section{Quantum Wire}

Considering a quantum rather than a classical wire, additional noise appears
due to the probabilistic nature of the charge transport (due to partitioning)
even in the absence of a fluctuating environment. Given a quantum wire with
$N$ channels characterized by their transmission eigenvalues $T^\gamma$,
$\gamma=1,\dots,N$, and biased by a voltage potential $V$, the characteristic
function of the full counting statistics is given by \cite{levitov:96}
\begin{equation}\label{eq:fcs_q}
  \log \chi^\text{q} (\lambda) = \frac{qV t}{2\pi\hbar} 
  \sum_\gamma \log[1+ (e^{i q \lambda}-1)T^\gamma],
\end{equation}
with $q$ the charge of the electron; this result is valid in the asymptotic
limit $qVt/\hbar \gg 1$, for low-temperatures $\vartheta\ll qV$, and with the
proviso that the energy-dependence of the transmission eigenvalues is
negligible in the energy interval set by the voltage.  The quantum nature of
the fermions leads to the noise
\begin{equation}\label{eq:noise_q}
  \langle \langle Q^2 \rangle\rangle^\text{q} = q \langle Q \rangle^\text{q} 
  \frac{\sum_\gamma T^\gamma (1-T^\gamma)}{\sum_\gamma T^\gamma}
\end{equation}
which disappears provided that all the channels are either closed $T^\gamma=0$
or completely open $T^\gamma=1$. Note that the noise in Eq.~(\ref{eq:noise_q})
is sub-Poissonian, i.e., the Fano factor $F=\langle\langle Q^2
\rangle\rangle^\text{q} /q \langle Q \rangle^\text{q}$ is less than 1.

Here, we are interested in the case where the quantum wire is capacitively
coupled to a two-level fluctuator such that the transmission eigenvalues
$T_\gamma$ change over time; note that we neglect a possible energy dependence
of the transmission eigenvalues, which corresponds to the fact that we assume
that the scattering center does not produce any time delay due to the
scattering event. The characteristic function $\log \chi^\text{q} (\lambda) =
\det \mathsf{Q}$ is given by the determinant of the matrix\cite{hassler:08}
\begin{equation}\label{eq:matrix}
  \mathsf{Q}_{k\gamma,k'\gamma'}= \langle \phi_{k \gamma}(t) | 
  e^{i \lambda q \mathcal{Q}_t}
  | \phi_{k'\gamma'}(t)\rangle
\end{equation}
with the counting operator $\mathcal{Q}_t = \int_I \! dx \, |x\rangle
\langle x|$ integrated over the interval $I = [0, v_\F t]$ and
\begin{equation}\label{eq:phi}
  \phi_{k\gamma}(x,t) =
  \begin{cases}
    e^{i k (x - v_\F t)} + r_\gamma(t+ x/v_\F) e^{-i k (x + v_\F t)} & x<0 \\
    \tau_\gamma( t- x/v_\F) e^{i k (x- v_\F t)}  & x>0
  \end{cases}
\end{equation}
the single-particle solution of the time-dependent Schr\"odinger equation
involving a scattering center at $x=0$ with time-dependent transmission
[reflection] amplitude $\tau_\gamma(t)$ [$r_\gamma(t)$]; note that we have
suppressed the transverse part of the wave function belonging to the channel
index $\gamma$. Equation (\ref{eq:phi}) is valid in the linear-spectrum
approximation, where $v_\F$ is the Fermi velocity and $0 \leq k \leq
k_\F$.\cite{linear_spectrum} As the matrix $\mathsf{Q}$ is block-diagonal in
$\gamma$ and constitutes a Toeplitz matrix with respect to the index $k$, its
determinant can be shown (using the technique discussed in Ref.\
\onlinecite{hassler:08}) to have the form
\begin{equation}\label{eq:logq}
  \log \chi^\text{q}(\lambda) = \frac{qV}{2\pi\hbar} \int_0^t \! dt'\,
  \sum_\gamma
  \log [1 + (e^{iq \lambda} -1) T_\gamma(t')]
\end{equation}
for long times $qV t/\hbar \gg 1$; the expression Eq.\ (\ref{eq:logq}) reduces
to Eq.\ (\ref{eq:fcs_q}) when the transmission probability does not change in
time.  We now add the classical two-level fluctuator to the system: Depending
on the state $x=x_{1,2}$ of the nearby two-level fluctuator, the quantum wire
is described by one of the two sets of transmission eigenvalues
$T^\gamma_{1,2}$. The total generating function $\chi^\text{qtl}(\lambda)$ is
an average over the individual contributions
\begin{align}\label{eq:fcs_qw}
  \chi^\text{qtl}(\lambda) &=  \langle \chi^\text{q}(\lambda) \rangle \nonumber\\
  &=
  \bigl\langle
  e^{(qV/2\pi\hbar) \int_0^t 
  dt' \sum_\gamma \log[1+ (e^{iq\lambda}-1)T^\gamma (t')]}
  \bigr\rangle ,
\end{align}
where the average $\langle \cdot \rangle$ is over the stochastic process
of the two-level fluctuator. Equation~(\ref{eq:fcs_qw}) can be calculated
explicitly with the method outlined in Sec.~\ref{sec:mapping}.  Indeed,
Eq.\ (\ref{eq:fcs}) goes over to Eq.\ (\ref{eq:fcs_qw}) via replacing $i
\lambda I(t)$ with $(qV/2 \pi \hbar) \sum_\gamma \log [ 1+ (e^{iq \lambda}
-1 ) T^\gamma (t)]$. In this mapping, the conductances $G_n$ are changed
to $(q/2 \pi i \hbar \lambda) \sum_\gamma \log [ 1+ (e^{iq \lambda} -1 )
T^\gamma_n]$ with $T^\gamma_n$ the transparency of the channel $\gamma$
when the two-level fluctuator is in the state $n$.  Inserting
\begin{equation}
 i \lambda \Delta g \to \mu = \frac{q V}{2\pi\hbar \Gamma} \sum_\gamma
  \log \Biggl[ \frac{1+ (e^{iq\lambda} -1) T^\gamma_2}{ 1+ (e^{iq\lambda}
  -1)
  T^\gamma_1} \Biggr]
\end{equation}
into Eq.\ (\ref{eq:fcs_large}) and using Eq.\ (\ref{eq:chi}), we obtain
\begin{align}\label{eq:fcs_gg_qw}
  \log \chi^\text{qtl}_\gg (\lambda) =&
  \sum_{n,\gamma} 
  P_n^\text{eq} \log \chi^q_n(\lambda)
  \\
  -&
  \frac{\Gamma t}{2} 
  \biggl[ 1+ \mu\, \Delta P^\text{eq} 
  -  \sqrt{ 1+ 2 \mu\, \Delta P^\text{eq} + \mu^2} \biggr];
  \nonumber
\end{align}
here, $\log \chi^q_n(\lambda)=(q Vt/2\pi\hbar) \sum_\gamma \log[ 1 +
(e^{iq \lambda} - 1) T^\gamma_n]$ is the characteristic function
for the FCS, Eq.~\eqref{eq:fcs_q}, dependent on the system's state
($n=1,2$) and $\gamma$ runs over the channel index.  The expression
\eqref{eq:fcs_gg_qw} coincides with the result obtained by Jordan and
Sukhorukov who considered the case of rare transitions ($\hbar\Gamma
\ll q V$), see Ref.~\onlinecite{jordan:04}.  In their work, Jordan and
Sukhorukov describe the transport of a conserved charge in a classical
bistable system where large charge fluctuations drive transitions between
stable states with different transport characteristics. The generic
fluctuations present in the system's stable states then combine with the
fluctuations of the bistable charge to generate the transport statistics of
the bistable system. While describing a very general situation, the specific
analysis in Ref.~\onlinecite{jordan:04} is limited to those cases where
the charge-switching events are rare on the time scale of the underlying
fluctuations in the stable states. The translation to our model may not
be obvious from the start and is done by choosing the process of charge
partitioning for generating the underlying fluctuations (with different
transmission rates $T_n^\gamma$, $n=1,2$ for the two stable states),
i.e., the generators $\log \chi^q_n(\lambda)$ correspond to the long time
generators $H_n t$ in Ref.~\onlinecite{jordan:04}. The non-linearity leading
to switching between stable states generates the transition probabilities
$\Gamma_{1,2}$; the latter are determined by an instanton trajectory and
have to be small in the case of Ref.~\onlinecite{jordan:04}. In our model
these rates are given as the basic input parameters $\gamma_{ij}$ defining
our two-level fluctuator. The condition of rare charge switching events
corresponds to the requirement that $\hbar \Gamma \ll qV$. In this limit
Eq.~(\ref{eq:fcs_gg_qw}) reduces to the result of the classical point
contact, Eq.~(\ref{eq:fcs_large}), with the conductances given by
\begin{equation}\label{eq:landauer}
  G_{1,2} = \frac{q^2}{2\pi \hbar} \sum_\gamma T^\gamma_{1,2}.
\end{equation}
We observe that the statistics is dominated by the fluctuations of the
impurity and the results derived in the previous sections remain valid in
the case of a quantum wire when the classical conductance is replaced by
the Landauer formula (\ref{eq:landauer}).

\begin{figure}[t]
  \centering
  \includegraphics[width=\linewidth]{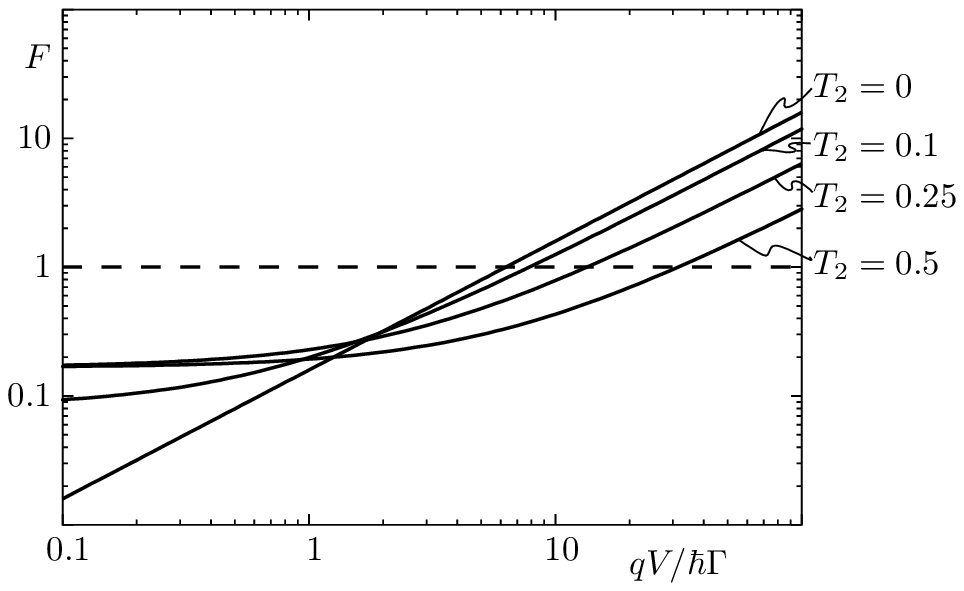}
  \caption{%
  Plot of the Fano factor $F$ as a function of the bias voltage $V$ for
  a single mode wire with $P^\text{eq}_1=P^\text{eq}_2=1/2$, $T_1=1$,
  and $T_2 = 0, 0.1, 0.3$, and 0.5 . The Fano factor starts of at a value
  $T_2(1-T_2)/(1+T_2) < 1$ for small voltages. In the opposite regime,
  it is given by $[(1-T_2)^2/(1+T_2)] qV/2\pi\hbar\Gamma$ which becomes
  superpoissonian for $V$ large enough.
  }\label{fig:crossover}
\end{figure}
However, our result Eq.~(\ref{eq:fcs_gg_qw}) is also valid in the opposite
regime $\hbar\Gamma\gg qV$, i.e., when the two-level fluctuator noise
acts on a timescale which is fast compared to the partitioning noise. Then
only the first term in (\ref{eq:fcs_gg_qw}) contributes and the cumulant
generating function is the average of the expressions (\ref{eq:fcs_q}) for
the quantum point contact, to be taken over the positions $x_{1,2}$ with
weights given by the probabilities $P^\text{eq}_{1,2}$. Having access to
both regimes, it is possible to study the crossover from classical noise
(due to the two-level fluctuator) to quantum-partitioning noise (due
to the quantum point contact).  To this end, we calculate the long-time
asymptotics of the first two moments of (\ref{eq:fcs_qw}),
\begin{equation}\label{eq:current_qw}
  \langle Q \rangle_\gg^\text{qtl} = V (P_1^\text{eq} G_1 + P_2^\text{eq}
  G_2 )\, t
\end{equation}
for the average charge and
\begin{multline}\label{eq:noise_qw}
  \langle\langle Q^2 \rangle\rangle_\gg^\text{qtl} = q V
  t\frac{q^2}{2\pi\hbar}\sum_{n,\gamma} P_n^\text{eq}  T^\gamma_n (1-
    T^\gamma_n) \\
 +2 P_1^\text{eq} P_2^\text{eq} \frac{V^2 (G_1 - G_2)^2 t}{\Gamma} 
\end{multline}
for the noise. The noise is simply given by the sum of the quantum
partitioning noise (first term) and the noise due to the dynamics of the
impurity (second term).  Note the crossover of the noise (\ref{eq:noise_qw})
from sub-Poissonian $F\leq 1$ for a fast fluctuator with $\hbar\Gamma \gg
qV$ [the first term in (\ref{eq:noise_qw}) dominates] to super-Poissonian
$F\geq 1$ when the fluctuator is slow $\hbar\Gamma \ll qV$ [the second
term in (\ref{eq:noise_qw}) dominates], provided that $G_1 \neq G_2$,
see Fig.~\ref{fig:crossover}.

\section{Experimental test}

It is difficult to experimentally confirm the crossover from sub-
to super-Poissonian noise as described in the previous section as the
quantum-partitioning noise is typically small and thus the noise of the
classical two-level fluctuator will dominate. As a promising setup, we
envision coupling a quantum point contact in GaAs/AlGaAs to a double dot,
e.g., in an InAs nanowire, which serves as tunable two-level fluctuator. Such
a system was studied recently, see Ref.~\onlinecite{kung:09}.

To observe the crossover, both the noise $S_\text{tl}(0)$ from the classical
two-level fluctuator and the quantum-partitioning noise $S_\text{qp}(0)$
(at zero frequency) have to dominate over the thermal Nyquist-Johnson
noise which is given by
\begin{equation}\label{eq:thermal_noise}
  S_\text{NJ}(0)  \approx  \frac{2 q^2 k_B T}{\pi\hbar} 
  \simeq 10^{-28} \,\text{A}^2
    \text{s}
\end{equation}
at $50\,$mK.  The quantum-partitioning noise [first term in
Eq.~\eqref{eq:noise_qw}] can be estimated as
\begin{equation}\label{eq:noise_qp}
S_\text{qp}(0) \approx \frac{q^3 V}{8\pi\hbar} \simeq V [\text{mV}] 10^{-27} 
\,\text{A}^2
  \text{s}
\end{equation}
when the system is tuned in the middle of a conductance step with $T\approx
1/2$.  Note that for bias voltages $V\geq 0.1\,$mV the quantum-partitioning
noise is larger than the thermal noise floor.

Tunneling of an electron between the two quantum dots with a rate $\Gamma$
leads to a change in the conductance. In Ref.~\onlinecite{kung:09}, this
change was of the order of $0.1\,q^2/\hbar$. However, in order to being
able to observe the crossover the capacitive coupling of the double dot to
the quantum point contact should be reduced to a level such that $G_1 -
G_2 \approx 0.001\,q^2/\hbar$.  This provides us with the estimate (using
$P_1^\text{eq}= P_2^\text{eq}=1/2$)
\begin{equation}\label{eq:noise_tl}
  S_\text{tl}(0) =
   \frac{V^2 (G_1 - G_2)^2}{\Gamma}
   \simeq  \frac{V^2 [\text{mV}^2]}{\Gamma [\text{MHz}]}
   10^{-26}\,\text{A}^2 \text{s}
 \end{equation}
for the zero-frequency noise due to the two-level fluctuator [second term
in Eq.~\eqref{eq:noise_qw}]. At a bias voltage $V\simeq 1\,$mV
 with a rate $\Gamma \simeq 100\,$MHz, the noise due to the fluctuator is
given by $S_\text{tl}(0)\simeq 10^{-28}\,\text{A}^2 \text{s}$
dominated by the quantum-partitioning noise with $S_\text{tl}(0)\simeq
10^{-27}\,\text{A}^2 \text{s}$. Note that in experiments rates of the order
of $10-100\,$kHz have been observed.\cite{hanson:07,kung:09} We expect
that rates in the 100$\,$MHz regime to be realistic due to the exponential
dependence of the tunneling rate on the potential barrier. Increasing
the bias voltage $V$ to values of a few mV the noise due to the classical
two-level fluctuator starts to dominate and the crossover as depicted in
Fig.~\ref{fig:crossover} can be observed.

\section{Conclusion}

We have determined the influence of a thermally driven two-level fluctuator
on a point contact through the calculation of the full counting statistics
of transported charge. Both, classical and quantum point contacts have been
considered and extensions to multiple fluctuators have been discussed. In
our analysis, we have made use of a mapping between correlation functions of
classical stochastic processes and simple time-ordered matrix products. For
the case of a quantum point contact, we have shown that the partitioning
noise and the noise due the two-level fluctuator add up and the noise crosses
over from sub- to super-Poissonian depending on the applied voltage bias.
To extend the present formalism to the case of current correlators at
finite frequencies, which provides additional insights into the dynamics
of the two-level fluctuator, is an interesting problem for future studies.

\acknowledgments

We acknowledge fruitful discussions with Misha Suslov and financial
support from the Pauli Center at ETH Zurich, the Swiss National Foundation,
the Russian Foundation for Basic Research (11-02-00744-a), the program
`Quantum physics of condensed matter' of the RAS, and the Computer Company
NIX (F1025/10-04).


\begin{thebibliography}{10}
\bibitem{linear_spectrum}
Note that the time-dependence of the scattering matrix promotes the incoming
  state to higher energies. In order that the linear spectrum approximation is
  still valid the time dependence of $T(t)$ should not be too abrupt.
\end{thebibliography}
\end{document}